\documentclass[twocolumn,pre,superscriptaddress,aps,footinbib]{revtex4}
\usepackage{url,psfrag}
\usepackage{dcolumn}
\usepackage{amsmath,amssymb}
\usepackage{bm}
\usepackage{hyperref}
\usepackage{float,epsfig,color}
\usepackage{times}
\usepackage[latin1]{inputenc}

\usepackage{graphicx}
\usepackage{amsmath}
\usepackage{amssymb}

\begin{document}

\begin{abstract}
Symmetries play a conspicuous role in the large-scale behavior of critical systems. In equilibrium they allow to classify asymptotics into different universality classes and, out of equilibrium, they sometimes emerge as collective properties which are not explicit in the ``bare'' interactions. Here we elucidate the emergence of an up-down symmetry in the asymptotic behavior of the stochastic scalar Burgers equation in one and two dimensions, manifested by the occurrence of Gaussian fluctuations even within the time regime controlled by nonlinearities. This robustness of Gaussian behavior contradicts naive expectations, due to the detailed relation ---including the lack of up-down symmetry--- between Burgers equation and the Kardar-Parisi-Zhang equation, which paradigmatically displays non-Gaussian fluctuations described by Tracy-Widom distributions. We reach our conclusions via a dynamic renormalization group study of the field statistics, confirmed by direct evaluation of the field probability distribution function from numerical simulations of the dynamical equation.
\end{abstract}

\title{Gaussian statistics as an emergent symmetry of the stochastic scalar Burgers equation}
\author{Enrique Rodr\'{\i}guez-Fern\'andez}
\email{enrodrig@math.uc3m.es}
\affiliation{Departamento de Matem\'aticas and Grupo Interdisciplinar de Sistemas Complejos (GISC)\\ Universidad Carlos III de Madrid, Avenida de la Universidad 30, 28911 Legan\'es, Spain}
\author{Rodolfo Cuerno}
\email{cuerno@math.uc3m.es}
\affiliation{Departamento de Matem\'aticas and Grupo Interdisciplinar de Sistemas Complejos (GISC)\\ Universidad Carlos III de Madrid, Avenida de la Universidad 30, 28911 Legan\'es, Spain}

\maketitle

\section{Introduction}

Spontaneous symmetry {\em breaking} is a basic notion in Physics underlying collective behavior in classical and quantum systems \cite{Chaikin95}. Among other important phenomena, it provides the mechanism for continuous phase transitions in equilibrium Statistical Mechanics, wherein the macroscopic state of a system shows a reduced symmetry compared with the microscopic interactions when temperature $T$ is below a certain threshold $T_c$. As is well known, the corresponding (critical) system is remarkably characterized by scale-invariant behavior right at $T=T_c$ \cite{Sethna06}. The converse situation of {\em emergent symmetries} occurs when the system symmetries {\em increase} for a decreasing $T$ \cite{Batista04}. This can occur even for non-equilibrium systems, whose large-scale behavior can display symmetries which are not explicit in the microscopic description. Recent examples include driven exciton-polariton condensates \cite{Sieberer13}, which give rise to novel dynamic universality classes beyond the standard classification of dynamical phase transitions \cite{Taeuber14}.

Actually, the generalization of the criticality concept to non-equilibrium conditions is proving itself a truly fruitful avenue to enlarge the domain of applicability of Statistical Physics, to e.g.\ socio-technological \cite{Castellano09} or living \cite{Munoz18} systems. In this process, an important conceptual role is being played by the elucidation of conditions for the generic occurrence of critical behavior without the need (in contrast with equilibrium systems) for parameter tuning, both in the presence or absence of a time-scale separation between external driving and system relaxation, termed self-organized criticality (SOC) \cite{Pruessner12} or generic scale invariance (GSI), respectively \cite{Grinstein95,Taeuber14}. A prime example for GSI is the Kardar-Parisi-Zhang (KPZ) equation for a scalar time-dependent field $h(\mathbf{r},t)$, namely,
\begin{eqnarray}
 & \partial_t h = \nu \nabla^2 h + (\lambda/2) (\nabla h)^2 + \eta, & \label{eq:kpz} \\
 & \langle \eta ({\bf r},t) \eta ({\bf r'},t') \rangle =2D\delta (\mathbf{r}-\mathbf{r}') \delta (t-t'), &
\label{eq:noise}
\end{eqnarray}
where $\nu, D >0$ and $\lambda$ are parameters, $\mathbf{r}\in\mathbb{R}^d$, and $\eta$ is {\em non-conserved}, zero-mean, uncorrelated Gaussian noise. Having been seminally put forward \cite{Kardar86} right at the crossroads among important domains of non-equilibrium phenomena ---like randomly stirred fluids, polymer dynamics in disordered media, and surface kinetic roughening---, the KPZ equation is recently being found to describe the universal behavior of a surprisingly wide range of strongly correlated systems \cite{Halpin-Healy15}, like bacterial range expansion \cite{Hallatschek07}, diffusion-limited growth \cite{Nicoli09}, turbulent liquid crystals \cite{Takeuchi12}, classical non-linear oscillators \cite{VanBeijeren12}, stochastic hydrodynamics \cite{Mendl13}, reaction-limited growth \cite{Alves13},
random geometry \cite{Santalla15}, superfluid exciton polaritons \cite{Altman15}, or incompressible polar active fluids \cite{Chen16}.

As elucidated analytically in one dimension (1D) \cite{Sasamoto10Amir11,Calabrese11} and numerically in 2D \cite{Halpin-Healy12,Oliveira13}, a remarkable trait of the KPZ universality class is the statistics of the fluctuations of the field $h$, which happens to
depend on global constraints on the dynamics, like the eventual time-dependence of the system size \cite{Halpin-Healy15,Corwin12}. The probability distribution function (PDF) for a suitably rescaled field is universal and of the celebrated Tracy-Widom (TW) family \cite{Corwin12}. In particular, the universal non-zero skewness for such PDF is interpreted \cite{Krug97} as reflecting a privileged direction for fluctuations in $h$ (e.g., a specific growth direction for the surface of a thin film \cite{Barabasi95,Kardar12}), as could be guessed by the lack of up-down symmetry of Eq.\ \eqref{eq:kpz} under $h\leftrightarrow -h$ \cite{Krug97}. Likewise, e.g.\ the nonlinear molecular-beam epitaxy (NLMBE) equation, a conserved-dynamics generalization of Eq.\ \eqref{eq:kpz} which also lacks the up-down symmetry \cite{Barabasi95,Krug97,Kardar12}, similarly displays non-zero skewness, even if the PDF does not belong to the TW family \cite{Carrasco16}.

For continuum models related with Eq.\ \eqref{eq:kpz}, which are frequently employed to explore critical dynamics far from equilibrium \cite{Kardar12}, we show in this article that the statistics of the evolving fluctuating field can differ from expectations based on straightforward analysis of the symmetries of its ``microscopic'' description. Indeed, taking the celebrated 1D Burgers equation \cite{Bec07} and its scalar 2D generalizations \cite{Hwa92,Vivo14} as representative cases, we find that Gaussian statistics are more robust than might have been expected when the large-scale behavior is controlled by a nonlinearity which breaks the up-down symmetry. Specifically, the stochastic scalar Burgers equation reads \cite{Forster77,Bertini94}
\begin{equation}\label{Burgers}
\partial_t \phi = \nu \partial_x^2 \phi + \lambda \phi \partial_x \phi + \eta,
\end{equation}
where $\eta$ is non-conserved noise exactly as in Eq.\ \eqref{eq:noise}. {\em In the absence of noise}, Eq.\ \eqref{Burgers} is obtained as the space-derivative of the deterministic Eq.\ \eqref{eq:kpz}, with $\phi=\partial_x h$ \cite{Kardar86}. The full stochastic Eq.\ \eqref{Burgers} can still be interpreted as a generalization of Eq.\ \eqref{eq:kpz} for a specific type of space-correlated noise, see e.g.\ \cite{Frey99}. We presently view Eq.\ \eqref{Burgers} as an instance of conserved dynamics with non-conserved noise (hence displaying GSI \cite{Taeuber14,Grinstein95}) which shares with Eq.\ \eqref{eq:kpz} the celebrated Galilean invariance \cite{Forster77} (i.e., it remains invariant under a Galilean change of coordinates) and the lack of symmetry under reflection in the field, $\phi\leftrightarrow -\phi$. Equation \eqref{Burgers} also shares the type of dynamics and noise, and the lack of up-down symmetry, with the NLMBE equation. However, as shown below, and in contrast with the cases of the latter and of the KPZ equation, the field statistics predicted by Eq.\ \eqref{Burgers} are Gaussian at large scales, an effective up-down symmetry emerging in its asymptotic nonlinear behavior. We reach this conclusion through a combined numerical and renormalization-group (RG) study which addresses the statistics of the physical field through its cumulants (analytically), and the full PDF (numerically).

Actually, Burgers equation, Eq.\ (\ref{Burgers}), is by itself another paradigm for non-equilibrium physics, appearing in many different contexts, like traffic models, cosmology, or turbulence, with different meanings for the field $\phi$, like vehicle density, mass density, or fluid velocity, respectively \cite{Bec07}. Moreover, the scalar Eq.\ (\ref{Burgers}) can be generalized to $d=2$ as, e.g.\ \cite{Vivo14}
\begin{equation}\label{Hwa92}
\partial_t \phi = \nu_x \ \partial_x^2 \phi + \nu_y \ \partial_y^2 \phi
+ \lambda_x \phi \partial_x \phi + \lambda_y \phi \partial_y \phi + \eta .
\end{equation}
The particular $\lambda_y=0$ case was originally introduced by Hwa and Kardar (HK) as a continuum model of avalanches in running sandpiles \cite{Hwa92} in the SOC context. We refer to the full Eq.\ \eqref{Hwa92} as the generalized Hwa-Kardar (gHK) equation.

\section{Generic scale invariance}

Both the KPZ and the stochastic Burgers equations exhibit GSI \cite{Taeuber14,Barabasi95,Grinstein95}: for arbitrary parameter values, the variance $W^2$ of the field grows with time $t$ as $W \sim t^{\beta}$ up to a saturation value $W_{sat} \sim L^{\alpha}$ at time $t_{sat} \sim L^z$, where $L$ is the lateral system size and $\beta=\alpha/z$. Universality classes occur, which are characterized by the values of the roughness or wandering exponent $\alpha$ (related with the fractal dimension of field configurations \cite{Barabasi95}) and of the dynamic exponent $z$ which characterizes the critical dynamics of these systems \cite{Kardar12,Taeuber14}, and by the statistics of fluctuations in the field, normalized as
\begin{equation}\label{chi}
    X (x,\Delta t,t_0) = \frac{\Delta \phi - \overline{\Delta \phi}}{(\Gamma \Delta t)^{\beta}},
\end{equation}
where $\Delta \phi (x,\Delta t,t_0)= \phi (x,t_0+\Delta t) - \phi (x,t_0)$, bar denotes space average, $\Gamma$ is a normalization constant \cite{Takeuchi13}, and $\Delta t \gg 1$ will be assumed. The statistical distribution of the fluctuations can differ before ($t_0=0, \Delta t \ll t_{sat}$) and after ($t_0 > t_{sat}$) saturation. For instance, for the 1D KPZ equation in band geometry the statistics are provided by the TW PDF for the largest eigenvalue of random matrices in the Gaussian Orthogonal Ensemble for $t_0=0,\Delta t \ll t_{sat}$ and by the Baik-Rains distribution for $t_0>t_{sat}$ and $\Delta t/t_0<1$ \cite{Takeuchi13,Halpin-Healy15}.

\begin{figure}[!t]
\begin{center}
\includegraphics[width=1.0 \columnwidth]{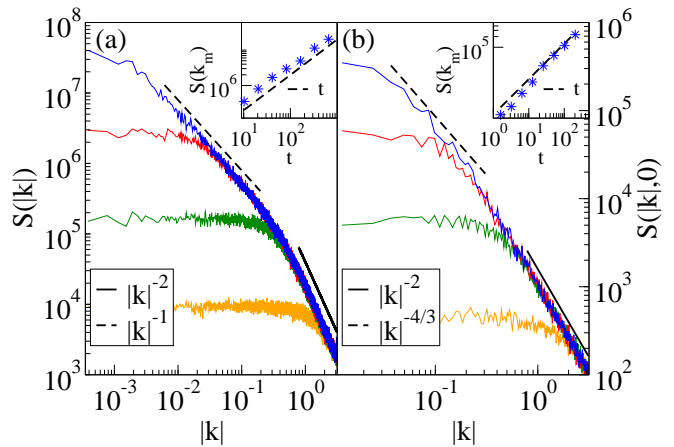}
\caption{$S(\mathbf{k},t)$ vs $k$ for increasing times [bottom to top, (a) $t=0.32,5.1,82,1300$; (b) $t=0.8,6.4,51,410$] for (a) Eq.\ (\ref{Burgers}), $L=2^{14}$, and (b) Eq.\ (\ref{Hwa92}) [cuts of $S(k_x,k_y)$ for $k_y=0$, $L=2^{9}$], for $\nu=\lambda=\nu_x=\lambda_x=\nu_y=\lambda_y=D=1$. Insets show $S(k_m,t)$ vs $t$. Straight lines have the indicated slopes, with dashed lines using the exponent values given in Table \ref{tablaexp}.
All units are arbitrary.}
\label{FactorEstructura}
\end{center}
\end{figure}

The scaling exponents of Eqs.\ (\ref{Burgers}) and (\ref{Hwa92}) have been investigated analytically \cite{Forster77,Hwa92,Frey99,Vivo14} and numerically \cite{Vivo12,Vivo14,Hayot97}, and are collected in Table \ref{tablaexp}. Note, HK scaling is anisotropic, hence the different exponent values along the $x$ and $y$ directions, while $\alpha_x/z_x=\alpha_y/z_y=\beta$ \cite{Vivo12}. As an illustration, Fig.\ \ref{FactorEstructura}(a) shows the time evolution of the structure factor for Eq.\ \eqref{Burgers}, $S(\mathbf{k},t)=\langle |\tilde{\phi}(\mathbf{k},t)|^2 \rangle$, where tilde is space Fourier transform, $\mathbf{k}$ is wave vector, and brackets are noise averages. Here and below, numerical simulations employ the pseudospectral method developed in \cite{Gallego11} for periodic systems.
As expected, for $|\mathbf{k}|$ larger than the inverse correlation length, power-law behavior ensues as $S(|\mathbf{k}|) \sim |\mathbf{k}|^{-(2\alpha+d)}$ \cite{Barabasi95,Krug97,Kardar12}. For Eq.\ \eqref{Burgers}, the system crosses over from linear behavior at short times and large $|\mathbf{k}|$ [the $|\mathbf{k}|$-dependent behavior of $S(|\mathbf{k}|)\sim |\mathbf{k}|^{-2}$ being induced by the linear term in the equation] to nonlinear behavior at long times and small $|\mathbf{k}|$, where $S(|\mathbf{k}|)\sim |\mathbf{k}|^{-1}$, inducing $\alpha=0$. In turn, $z=1$ is implied [inset of Fig.\ \ref{FactorEstructura}(a)] by the $S(k_m,t) \sim t^{(2\alpha+d)/z}$ scaling at the smallest wave vector in the system, $k_m=2\pi/L$ \cite{Barabasi95,Krug97,Kardar12}. The $\alpha=0$, $z=1$ values thus obtained for the asymptotics of Eq.\ \eqref{Burgers} equal, incidentally, those of a {\em linear, non-local} continuum model that describes diffusion-limited erosion (DLE) \cite{Krug91,Krug97}.
\begin{table}[!tb]
\begin{center}
\begin{tabular}{|c|c|c|c|c|c|}
\hline
Equation & $\alpha_x$ & $\alpha_y$ & $ z_x $ & $ z_y $ & $\beta$ \\
\hline \hline
1D Burgers & $0$ & not defined & $1$ & not defined & $0$ \\ \hline
Hwa-Kardar & $-1/5$ & $-1/3$ & $6/5$ & $2$ & $-1/6$ \\ \hline
g Hwa-Kardar & $-1/3$ & $-1/3$ & $4/3$ & $4/3$ & $-1/4$ \\ \hline
\end{tabular}
\caption{Scaling exponents for Eqs.\ \eqref{Burgers}, \eqref{Hwa92} \cite{Forster77,Hwa92,Vivo12,Frey99,Vivo14,Hayot97}.}
\label{tablaexp}
\end{center}
\end{table}
The numerical data in Fig.\ \ref{FactorEstructura}(b) can be similarly discussed to justify the corresponding entries in Table \ref{tablaexp} for the asymptotic behavior of the gHK equation in $d=2$, see likewise \cite{Vivo12} for HK.

\section{Dynamical Renormalization Group analysis}

While $\alpha\leq 0$ as in Table \ref{tablaexp} usually indicates that $d$ is at or above the upper-critical dimension $d_c$ \cite{Chaikin95,Sethna06}, for Eqs.\ (\ref{Burgers}) and (\ref{Hwa92}) $d_c=4$ has been demonstrated \cite{Forster77,Hwa92}. Specifically, for Eq.\ \eqref{Burgers} $\alpha=0$ suggests the validity of the Gaussian approximation, while asymptotics is nonlinear. Hence, it is interesting to study the scaling behavior of this equation in detail. We resort to a dynamic RG (DRG) analysis, which has been successfully employed in this context \cite{Forster77,Yakhot86}, being based on an iterative solution of Eq.\ (\ref{Burgers}) in Fourier space,
where it reads
\begin{align}
 & G(k,\omega) \hat{\eta} = G_0(k,\omega) \hat{\eta} + \lambda G_0(k,\omega) \times \label{Fourier}\\
 & \times ({\rm i} k) \int_{-\infty}^{\infty} \frac{d\Omega}{2\pi} \int_{-\Lambda_0}^{\Lambda_0} \frac{dq}{2\pi} \ G(q,\Omega) \hat{\eta} \  G(k-q,\omega-\Omega) \hat{\eta}, \nonumber
\end{align}
with $G_0(k,\omega)=(-{\rm i}\omega + \nu k^2)^{-1}$, $G(k,\omega)=(-{\rm i}\omega + \tilde{\nu}(k) k^2)^{-1}$, $G(k,\omega) \hat{\eta} = \hat{\phi}(k,\omega)$, hat is space-time Fourier transform, $k$ is wave-number, $\omega$ is time frequency, and $\rm{i}$ is the imaginary unit. After coarse-graining and rescaling, the one-loop DRG flow for parameters $\nu$, $\lambda$, and $D$ reads \citep{Forster77} (see Appendix \ref{Ap:Prop} for details)
\begin{eqnarray}\label{FlujoRenorm}
    \frac{d \nu}{d\ell} & = & \nu \left( z - 2 + \frac{6}{\pi} \frac{\lambda^2 D}{\nu^3} \right),
    \\
    \frac{d \lambda}{d\ell} & = & \lambda(\alpha + z - 1), \quad     \frac{d D}{d\ell} = D(z-2\alpha-1).
\label{renormLD}
\end{eqnarray}
Requesting scale invariance at a non-linear ($\lambda\neq0$) critical point leads to $\alpha+z=1$, associated with the Galilean invariance of Eq.\ \eqref{Burgers}. Non-trivial fluctuations ($D\neq0$) require non-renormalization of the noise, leading to hyperscaling \cite{Forster77,Barabasi95}, $2\alpha+d=z$ (with $d=1$), due to the fact that dynamics are conserved but the noise is not \cite{Grinstein95}. These two scaling relations are believed to hold at any order in the loop expansion \cite{Forster77,Hwa92}. They provide an equation set for $\alpha$ and $z$ whose unique solution in $d=1$ ($2$) is the Burgers (gHK) row in Table \ref{tablaexp} \cite{VivoN}.

\subsection{DRG evaluation of cumulants}

Having determined the scaling exponents, we henceforth perform a partial RG transformation only, which omits the rescaling step. This allows to make explicit the scale-dependence of the equation parameters, as proposed in \cite{Yakhot86}. While $\lambda$ and $D$ do not renormalize and are thus scale-independent, the coarse-grained linear coefficient $\tilde{\nu}(k)$ depends on wave vector as
$
    \tilde{\nu}(k) \simeq (6D\lambda^2/\pi)^{1/3} |k|^{-1} 
$
(see Appendix \ref{Ap:Prop} for details). We exploit this fact to estimate by DRG the cumulants of the statistical distribution of $\phi$, following the methodology successfully employed for the KPZ \cite{Singha14,Singha15,Singha16b} and NLMBE \cite{Singha16} equations. Thus, the $n$-th cumulant reads
\begin{equation}
\langle \phi^n \rangle_c = \int_{\mathbb{R}^{2(n-1)}} G(k_n,\omega_n) L_n  \prod_{j=1}^{n-1} \frac{dk_j d\omega_j}{(2\pi)^2} G(k_j,\omega_j),
\end{equation}
where $k_n=-\sum_{j=1}^{n-1}k_j$, $\omega_n=-\sum_{j=1}^{n-1}\omega_j$, and the function $L_n$ needs to be perturbatively computed (see Appendix \ref{Ap:Cum} for details). Finally,
\begin{eqnarray}
\langle \phi^n \rangle_c & = &  \frac{A}{D^{n-1} \lambda^{3n-2}} \int_{\mathbb{R}^{2(n-1)}} G(k_n,\omega_n) k_n \nonumber \\
 & \times & \prod_{i=1}^{n-1} \frac{dk_i d\omega_i}{(2\pi)^2} k_i G(k_i,\omega_i) |G(k_i,\omega_i)|^2,
\label{CumulanteFinal}
\end{eqnarray}
where $A=\pi^{2n-\frac{1}{2}} {\rm i}^n \Gamma (n-\frac{1}{2}) K/[n!(n-1) 3^n 2^{2n-1}]$. Integration of Eq.\ (\ref{CumulanteFinal}) for $n=2$ yields the variance of $\phi$,
\begin{equation}
    \langle \phi^2 \rangle_c =\frac{1}{(3072\pi^2)^{1/3}} \left( \frac{D}{\lambda} \right)^{2/3} \int_{\mathbb{R}} \frac{dk}{|k|},
\end{equation}
whose logarithmic divergence ($\sim \ln L$) agrees with the expected value of the roughness exponent, $\alpha=0$ \cite{Krug91,Krug97}.

For odd cumulants (odd $n$), after integration in $\omega_1,...,\omega_{n-1}$, the integrand of Eq.\ (\ref{CumulanteFinal}) equals
$
    k_n g(k_1,...,k_{n}) \prod_{i=1}^{n-1} k_i,
$
where all $k_i$ in $g(\cdot)$ are to be taken in absolute value. Now, this expression is antisymmetric under the transformation $k_i \mapsto -k_i$, which maps the semispace
$ S_+ = \{ (k_1,...,k_{n-1}) \in \mathbb{R}^{n-1} | \sum_{i=1}^{n-1} k_i > 0 \} $
into
$ S_- = \{ (k_1,...,k_{n-1}) \in \mathbb{R}^{n-1} | \sum_{i=1}^{n-1} k_i < 0 \} $.
Hence, the integral over the full $\mathbb{R}^{n-1}$ cancels exactly. Thus, all the odd cumulants of the $\phi$ distribution are zero, leading to a symmetric PDF \citep{Gardiner09,Demostracion}.

\begin{figure}[!t]
\begin{center}
\includegraphics[width=1.0 \columnwidth]{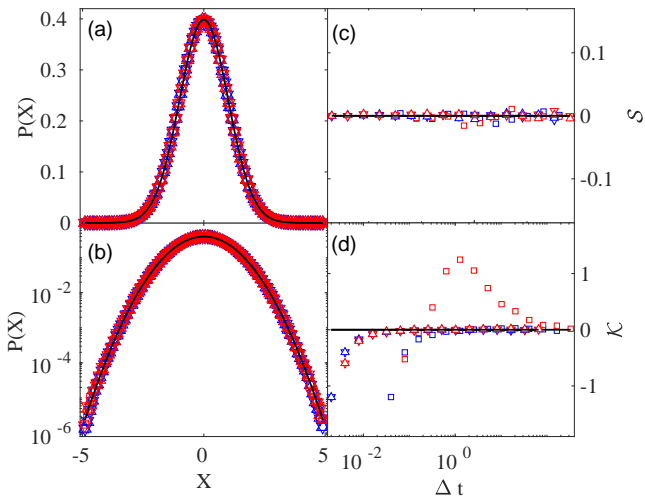}
\caption{Fluctuation histogram [(a), (b)] from direct simulations of the Burgers ($\square$), HK ($\triangle$), and gHK ($\triangledown$) equations, Eqs.\ (\ref{Burgers}) and (\ref{Hwa92}). Here, $X$ is as in Eq.\ (\ref{chi}), blue \ (red) denotes $t_0=0,\Delta t \ll t_{sat}$ ($t_0>t_{sat}$), with $\Delta t\gg 1$ as discussed after Eq.\ \eqref{chi}. The solid line shows the exact Gaussian PDF. Full time evolution of the (c) skewness, $\mathcal{S}$ and (d) excess kurtosis, $\mathcal{K}$. Symbols as in (a,b). Convergence to Gaussian (zero) values occurs in all cases.}
\label{Histograma}
\end{center}
\end{figure}

The fourth cumulant is estimated by means of analytical integration in time frequencies $\omega_i$ and numerical integration in wavenumbers $k_j$. Specifically, using a lower cut-off for the latter, $\mu \propto 1/L$ in the $10^{-3}-10^{-8}$ range, the integral
(to simplify the notation, we drop $\omega$-dependences in $G$ and $L_4$)
\begin{align}
    & \iiint_{\mathbb{R}^3} \prod_{i=1}^3 \frac{d \omega_i}{2\pi}
    \iiint_{{\scriptscriptstyle \mu \leq |k_j| \leq \Lambda_0}}
    \prod_{j=1}^3 \frac{dk_j}{2\pi} G(k_1) G(k_2) G(k_3) \nonumber \\
    & \times G(-k_1-k_2-k_3) L_4(k_1,k_2,k_3) \label{eq:12}
\end{align}
diverges as $\langle \phi^4 \rangle_c \sim (\ln L)^{1.05}$ [see Appendix \ref{Ap:Kurt} for details]. Thus, the excess kurtosis of the distribution, $\mathcal{K}=\langle \phi^4 \rangle_c / \langle \phi^2 \rangle_c^{2}$, vanishes for increasing system size ($L \to \infty$).

\section{Direct numerical simulations}

With null odd cumulants and an excess kurtosis which decreases for increasing $L$, these analytical results indeed suggest Gaussian statistics for the field fluctuations in the stochastic Burgers equation \eqref{Burgers}. The exact cancellation of the odd cumulants is particularly remarkable, in view of the lack of up-down symmetry in the equation. Given the approximations made in our one-loop DRG analysis, we have carried out direct numerical simulations of the Burgers, the HK, and the gHK equations, in order to explicitly compute the full PDF in each case. Histograms have been constructed for times both in the nonlinear growth regime ($t_0=0,\Delta t \ll t_{sat}$) and after saturation to steady state ($t_0>t_{sat}$), using $L=2^{20}$ for Burgers and $L=2^{10}$ for the HK and gHK equations; other parameters are as in Fig.\ \ref{FactorEstructura}. In all cases the PDF is Gaussian to a high precision, compare the symbols in Fig.\ \ref{Histograma} with the exact Gaussian form (solid line). More quantitatively, Fig.\ \ref{Histograma} also shows the time evolution of the skewness $\mathcal{S}=\langle \phi^3 \rangle_c / \langle \phi^2 \rangle_c^{3/2}$ and excess kurtosis $\mathcal{K}$. While $\mathcal{S}(\Delta t)$ remains essentially null in all cases for Eqs.\ \eqref{Burgers} and \eqref{Hwa92}, convergence of $\mathcal{K}$ to zero requires sufficiently large $\Delta t$, specially for Eq.\ \eqref{Burgers}. All this supports our conclusions from the DRG analysis of the stochastic Burgers equation.

\section{Discussion}

Let us note that, since the scaling exponents of Eqs.\ (\ref{Burgers}) and (\ref{Hwa92}) fulfill hyperscaling, as does any linear model \cite{Krug97}, an interesting consequence of our results is that evolution equations can be formulated which share with Eqs.\ (\ref{Burgers}) and (\ref{Hwa92}) both, the exponent values and the Gaussian statistics, but which are linear (thus, up-down symmetric)! Namely, the Gaussian approximation becomes exact. Indeed, by writing
\begin{equation}\label{Proxy}
\partial_t \tilde{\phi} = \big( - \sum_{i=1}^{d} |k_i|^{z_d} \big) \, \tilde{\phi} + \tilde{\eta},
\end{equation}
the choice $z_1=1$ in $d=1$ (the continuum DLE model \cite{Krug91,Krug97}) yields the asymptotic behavior of Eq.\ (\ref{Burgers}), while in $d=2$, choosing $z_2=4/3$ provides the exponents and Gaussian statistics of the gHK equation, and similarly for the HK model using $z_2=6/5$ and dropping the $k_y$-dependence \cite{Vivo12}.

\begin{figure}[!t]
\begin{center}
\includegraphics[width=1 \columnwidth]{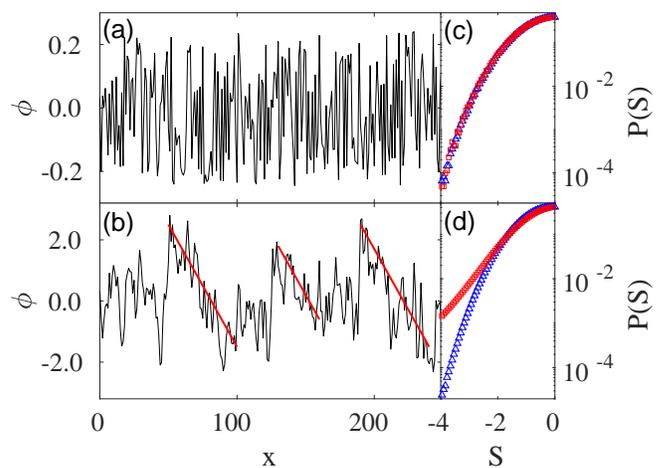}
\caption{Profiles described by Eq.\ \eqref{Burgers} in the linear (a) and nonlinear (b) regimes for parameters as in Fig.\ \ref{FactorEstructura}. Parallel straight lines are guides to identify sawtooth-like patches. The slope histogram for time as in (a) [(b)] appears in (c) [(d)], where $S=(\partial_x \phi - \overline{\partial_x \phi})/{\rm std}(\partial_x \phi)$ is normalized slope. The $S>0$ tail (red squares) is reflected to facilitate comparison with the $S<0$ (blue triangles) data.}
\label{Morfology}
\end{center}
\end{figure}

Focusing on the stochastic Burgers equation, it is the nonlinear term $\phi\partial_x\phi$ which is responsible for the up-down asymmetry, and it indeed plays an essential role in the nontrivial behavior described above. Actually, we can rationalize the emergence of the up-down symmetry in the asymptotic nonlinear regime by considering the effect of this term when isolated, i.e.\ for the inviscid deterministic Burgers equation, whose solutions are know analytically \cite{Bec07,Burgers74}. Note that this nonlinearity also breaks the left-right ($x \leftrightarrow -x$) symmetry in the system and indeed, as is well known, it generically induces sawtooth-like profiles \cite{Bec07,Burgers74,Bendaas18}, which notably are symmetric around their mean under a {\em combined} $(x,\phi) \leftrightarrow (-x,-\phi)$ reflection. Analogous behavior also occurs in the full stochastic Burgers equation, becoming even apparent to the naked eye in the asymptotic regime. It is illustrated in Fig.\ \ref{Morfology}, where typical morphologies are shown in the linear and nonlinear regimes. For the latter, the parallel straight lines allow to identify portions of the profile which are ``noisy sawtooths". Quantitative confirmation is provided by the slope histogram $P(S)$, obtained for the corresponding linear and nonlinear regimes and also given in Fig.\ \ref{Morfology}. While the distribution of slopes is symmetric for times dominated by the linear term in Eq.\ \eqref{Burgers}, the histogram becomes non-symmetric in the nonlinear regime, large positive slopes being much more frequent than before due to the appearance of the abrupt jumps in the $\phi$ values that can be seen in Fig.\ \ref{Morfology}(b), reminiscent of deterministic sawthooths. Thus, we believe that the asymptotic emergence of the up-down symmetry in Eq.\ \eqref{Burgers} can be traced back to the deterministic form of solutions induced by its nonlinearity, this mechanism being also operative in the HK and gHK equations. However, the competition with noise remains far from trivial in these systems, whose solutions, analogously to the KPZ case \cite{Kardar86}, differ quite strongly with those of their deterministic counterparts. 

\section{Summary and Conclusions} 

In summary, we have obtained that, although the asymptotic behavior of the universality class of the Burgers equation with non-conserved noise in $d=1, 2$ is controlled by nonlinear terms that break the up-down symmetry, the statistics are nonetheless Gaussian. This remains true under strongly-anisotropic perturbations, e.g.\ by setting $\lambda_y=0$ in the gHK equation to obtain the HK equation, with different exponents but still Gaussian statistics. Our result is in spite of the close relation of Eqs.\ \eqref{Burgers} and \eqref{Hwa92} with the KPZ equation, whose statistics are paradigmatically non-Gaussian, and contrasts with the non-zero skewness of the NLMBE equation too \cite{Carrasco16}.

Overall, Gaussian statistics can hence emerge for suitable systems whose bare interactions break the symmetries that one might naively associate with the former, at least when such symmetries are broken as in the KPZ case. We hope that our results may aid in the challenge of fully understanding fluctuations in spatially-extended systems far from equilibrium.

\begin{acknowledgments}
We acknowledge valuable comments by M.\ Castro, J.\ Krug, and P.\ Rodr\'{\i}guez-L\'opez, and funding by Ministerio de Econom\'{\i}a y Competitividad, Agencia Estatal de Investigaci\'on, and Fondo Europeo de Desarrollo Regional (Spain and European Union) through grant No.\ FIS2015-66020-C2-1-P. E.\ R.-F.\ acknowledges financial support by Ministerio de Educaci\'on, Cultura y Deporte (Spain) through Formaci\'on del Profesorado Universitario scolarship No.\ FPU16/06304.
\end{acknowledgments}

\appendix

\section{DRG analysis of the 1D stochastic Burgers equation}

\subsection{Propagator renormalization}\label{Ap:Prop}

Working within the one-loop approximation, the renormalized propagator $G(k,\omega)$ of the stochastic Burgers equation reads \cite{Forster77,Barabasi95}, to lowest order in $\lambda$,
\begin{equation}\label{General}
    G(k,\omega)=G_0(k,\omega)+ G_0^2(k,\omega) \Sigma_0,
\end{equation}
the Feynman representation of this equation being depicted in Fig.\ \ref{FeynmanNu}(a). Hence, $\Sigma_0$ is computed as
$$
\Sigma_0 = 4 \lambda^2  \int^> \frac{dq}{2\pi} {\rm i} k \ {\rm i} (k-q) \int_{-\infty}^{\infty}  \frac{d\Omega}{2\pi}
$$
\begin{equation}
    \times  |G_0(q,\Omega)|^2 G_0 (k-q,\omega-\Omega).
\end{equation}
After integration in $\Omega$, taking the long-time limit $\omega \rightarrow 0$,
\begin{eqnarray}
\Sigma_0 &=& - 4 \lambda^2  \int^> \frac{dq}{2\pi} \frac{Dk(k-q)}{\nu^2 q^2 (k^2-2kq+2q^2)} 
\nonumber \\
 & = & - 4 \lambda^2  \int^> \frac{dq}{2\pi} \left[ \frac{3 D k^2}{4 \nu^2 q^4} \right] + \mathcal{O}(k^3) 
\nonumber \\
 & = & \frac{6D\lambda^2 k^2}{\nu^2 \pi \Lambda^3(\ell)}(e^{-3\ell}-1)+\mathcal{O}(k^3).
\end{eqnarray}
As we are interested in the large-scale behavior, only the lowest order in $k$ will be considered. Now, after identifying $k=\Lambda_0 e^{-\ell}$, we can compute the renormalized $\tilde{\nu}(\ell)$ from Eq.\ (\ref{General}) in the long-time limit, as $G(k,0)=[\tilde{\nu}(k) k^2]^{-1}$ \cite{Yakhot86},
$$
    \frac{1}{\tilde{\nu}(\ell) k^2} = \frac{1}{\nu k^2} + \frac{1}{\nu^2 k^4} \frac{6D\lambda^2 k^2}{\nu^2 \pi \Lambda^3(\ell)}(e^{-3\ell}-1)
$$
\begin{eqnarray*}
 & \Rightarrow & 
\frac{1}{\tilde{\nu}(\ell)} = \frac{1}{\nu} \left[ 1 + \frac{6D\lambda^2}{\nu^3 \pi \Lambda^3(\ell)}(e^{-3\ell}-1) \right] \nonumber
\\
 & \Rightarrow & \tilde{\nu}(\ell) = \nu \left[ 1-\frac{6D\lambda^2}{\nu^3 \pi \Lambda^3(\ell)}(e^{-3\ell}-1) \right].
\end{eqnarray*}
This equation can be rewritten as a differential flow, thus
\begin{equation}
    \frac{d \tilde{\nu}}{d \ell} =  \frac{18D\lambda^2}{\nu^2 \pi \Lambda^3(\ell)},
\end{equation}
whose solution for the initial condition $\tilde{\nu}(0)=\nu$ is
\begin{equation}
    \tilde{\nu} (\ell ) = \left[\frac{\nu^3}{3} + \frac{6D\lambda^2}{\nu^3 \pi \Lambda^3_0} (e^{3\ell}-1) \right]^{1/3}.
\end{equation}
In the large-scale limit when $\ell \gg 1$, the renormalized coefficient $\tilde{\nu}$ scales with the wavenumber $k$ as \cite{Forster77}
\begin{equation}\label{RenorNu}
   \tilde{\nu}(k) \sim \left( \frac{6D\lambda^2}{\nu^3 \pi} \right)^{1/3} |k|^{-1}.
\end{equation}
This equation is explicitly given in the main text.

\subsection{Cumulants}\label{Ap:Cum}

\begin{figure}
\begin{center}
\includegraphics[width=1 \columnwidth]{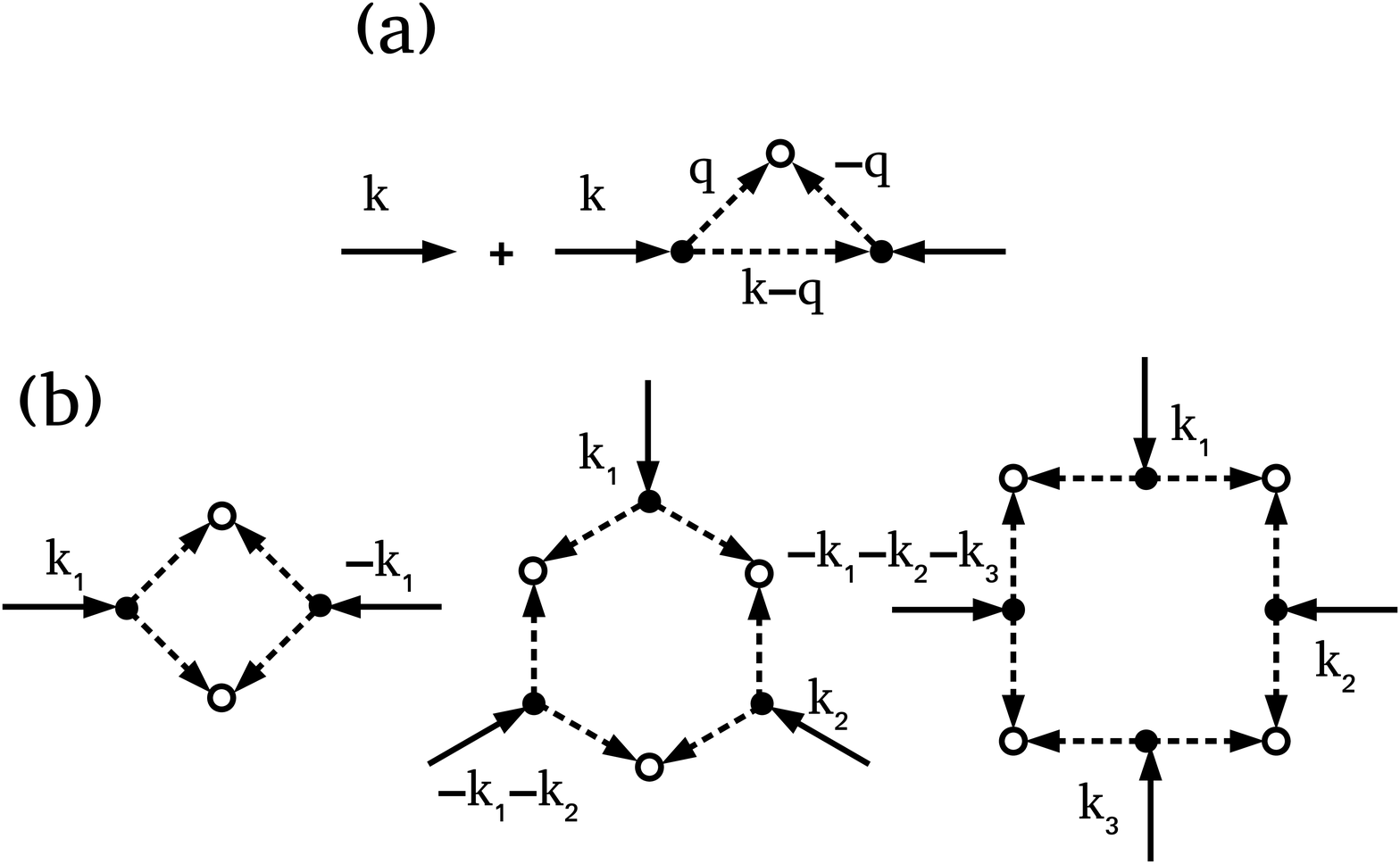} 
\caption{Feynman diagrams representing (a) Eq.\ \eqref{General} and (b) the lowest-order corrections to the cumulants, $L_{n,1}$, for $n=2,3,4$, left to right. Bare propagator factors $G_0(q)$ evaluated for low [$|q|<\Lambda(\ell)$] and high [$\Lambda(\ell)<|q|<\Lambda_0$] wave vectors correspond to solid and dashed lines, respectively. Noise contractions (convolution products) are represented by open (filled) disks.}
\label{FeynmanNu}
\end{center}
\end{figure}

The $n$-th cumulant of the field fluctuations reads, for the 1D stochastic Burgers equation,
\begin{equation}
\langle \phi^n \rangle_c = \int_{\mathbb{R}^{2(n-1)}} G(k_n,\omega_n) L_n  \prod_{j=1}^{n-1} \frac{dk_j d\omega_j}{(2\pi)^2} G(k_j,\omega_j),
\end{equation}
where $k_n=-\sum_{j=1}^{n-1}k_j$, $\omega_n=-\sum_{j=1}^{n-1}\omega_j$. The function $L_n$ is perturbatively computed to one loop order \cite{Singha14,Singha15,Singha16b,Singha16} as
\begin{equation}{}
    L_n=(2D) \delta_{n,2} +L_{n,1},
\end{equation}
where $
    L_{n,1} = K (2D)^n \lambda^n {\rm i}^n k_n \prod_{j=1}^{n-1} k_j l_{n,1} $
\ is the lowest-order correction in the Feynman expansion of the cumulants and  $K=(2n-2)!!$ is a combinatorial factor (number of different fully-connected diagrams). Diagrammatic representations for $L_{2,1}$, $L_{3,1}$, and $L_{4,1}$ correspond to the amputated parts of the diagrams shown in Fig.\ \ref{FeynmanNu}(b).

As we are interested in the $(k_i,\omega_i) \rightarrow (0,0)$ limit,
\begin{equation}{}
    l_{n,1}=\int_{-\infty}^{\infty} \frac{d\Omega}{2\pi} \int^> \frac{dq}{2\pi} |G_0(q,\Omega)|^{2n},
\end{equation}
where the integration domain in $\int^>$ is the region $\{ q\in \mathbb{R} | \, \Lambda(\ell)=\Lambda_0 e^{-\ell} <|q|<\Lambda_0 \}$.
After integration, and substituting $\nu \to \tilde{\nu}(\Lambda)$,
\begin{equation}{}
    l_{n,1} = \frac{\sqrt{\pi} \Gamma(n-\frac{1}{2})}{n! \left( \frac{6D\lambda^2}{\pi} \right)^{2n-1}} \frac{2\big(1-e^{-(4n-3) \ell}\big)}{(4n-3) \Lambda^{2n-2}(\ell)}.
\end{equation}
We rewrite this equation in differential form as
\begin{equation}
    \frac{dl_{n,1}}{d\ell}=\frac{\sqrt{\pi} \Gamma(n-\frac{1}{2})}{n! \left( \frac{6D\lambda^2}{\pi} \right)^{2n-1}} \frac{2}{\Lambda^{2n-2}(\ell)},
\end{equation}
whose solutions for large $\ell$ become
\begin{equation}
    l_{n,1}(\ell) \simeq \frac{\sqrt{\pi} \Gamma(n-\frac{1}{2})}{n! \left( \frac{6D\lambda^2}{\pi} \right)^{2n-1} (n-1)} \frac{1}{\Lambda^{2(n-1)}(\ell)}.
\end{equation}
Due to the symmetry among $k_1,\ldots,k_{n-1}$, we take \cite{Singha14,Singha15,Singha16b,Singha16}
\begin{equation}
    l_{n,1}(k)=\frac{\sqrt{\pi} \Gamma(n-\frac{1}{2})}{n! \left( \frac{6D\lambda^2}{\pi} \right)^{2n-1} (n-1)}
    \prod_{i=1}^{n-1} \frac{1}{k_i^2}.
\end{equation}
For $n > 2$, as $|G(k,\omega)|=|k|^{-z}f(\omega/|k|^z)$ and $z=1$, where $f$ is a scaling function [$f(u) \rightarrow 1$ for $u \rightarrow 0$],
$k_i^{-2} \simeq |G(k_i,\omega_i)|^2$. Finally,
\begin{eqnarray}
\langle \phi^n \rangle_c & = & \frac{A}{D^{n-1} \lambda^{3n-2}} \int_{\mathbb{R}^{2(n-1)}} G(k_n,\omega_n) k_n \nonumber \\
 & \times & \prod_{i=1}^{n-1} \frac{dk_i d\omega_i}{(2\pi)^2} k_i G(k_i,\omega_i) |G(k_i,\omega_i)|^2,
\end{eqnarray}
where $A=\pi^{2n-\frac{1}{2}} {\rm i}^n \Gamma (n-\frac{1}{2}) K/[n!(n-1) 3^n 2^{2n-1}]$. This is Eq.\ \eqref{CumulanteFinal} of the main text.

\subsection{Kurtosis scaling with system size}\label{Ap:Kurt}

\begin{figure}[!b]
\begin{center}
\includegraphics[width=1.0 \columnwidth]{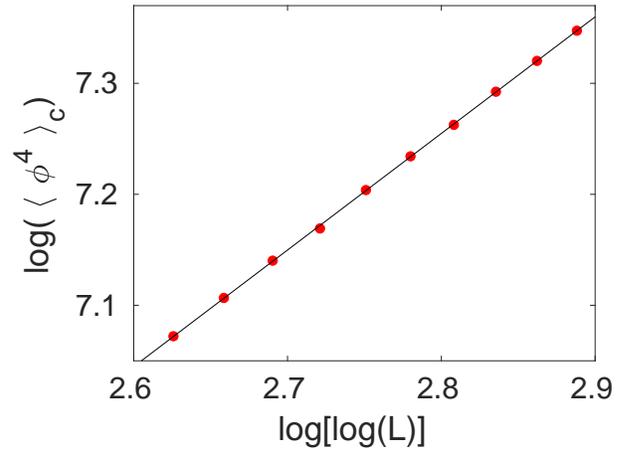} 
\caption{Numerical computation of the fourth cumulant in the $[k_1,k_2,k_3] \in [1/L,L]^3$ region, for different values of $L$ (symbols). The solid line shows a linear fit of the numerical data, and corresponds to the straight line $y=1.05x+4.314$, hence $\langle \phi^4 \rangle_c \sim (\ln L)^{1.05}$.}
\label{FigW4}
\end{center}
\end{figure}

The fourth cumulant of the fluctuation distribution has been estimated for different values of the system size $L$ by means of analytical integration in $\omega_1,\omega_2,\omega_3$ and numerical integration in $k_1,k_2,k_3$. Parameters have been chosen so as to make $A=1$ and $6D\lambda^2/\pi=1$. Integration limits in $k_1,k_2,k_3$ of the form $[1/L,L]$ have been taken for different values of $L$ in order to characterize the divergence of the integral with $L$. The conclusion is that $\langle \phi^4 \rangle_c \sim (\ln L)^{1.05}$, see Fig.\ \ref{FigW4}, a result which is employed after Eq.\ \eqref{eq:12} of the main text.


\begin{thebibliography}{99}
\bibitem{VivoN}
The case of the gHK equation is discussed in \cite{Vivo14}.
\bibitem{Demostracion} The PDF of the stochastic variable $X$ is $P(X)=\mathcal{F}^{-1}\{\exp [\sum_{n=1}^{\infty} ({\rm i}s)^n C_n (X)/n!]\}$, where $C_n (X)$ are the cumulants, $\mathcal{F}$ is Fourier transform and $s$ is conjugate to $X$. As ${\cal F}$ preserves the parity of a function, if all the odd-order cumulants are zero, the function of $s$ is even and $P(X)$ is symmetric.
\end{thebibliography}
\end{document}